\documentclass[12pt,preprint]{aastex}
\usepackage[dvips]{color}
\usepackage{graphicx}

\def\msol{\hbox{\kern 0.20em $M_\odot$}}
\def\lsol{\hbox{\kern 0.20em $L_\odot$}}
\def\rsol{\hbox{\kern 0.20em $R_\odot$}}
\def\sr{\hbox{\kern 0.20em sr}}
\def\srmu{\hbox{\kern 0.20em sr$^{-1}$}}
 
\def\g{\hbox{\kern 0.20em g}}
\def\gmu{\hbox{\kern 0.20em g$^{-1}$}}
\def\kg{\hbox{\kern 0.20em kg}}
\def\pc{\hbox{\kern 0.20em pc}}
 
\def\mum{\hbox{\kern 0.20em $\mu$m}}
\def\mumd{\hbox{\kern 0.20em $\mu$m$^{-2}$}}
\def\cm{\hbox{\kern 0.20em cm}}
\def\m{\hbox{\kern 0.20em m}}
\def\km{\hbox{\kern 0.20em km}}
\def\nm{\hbox{\kern 0.20em nm}}
 
\def\s{\hbox{\kern 0.20em s}}
\def\h{\hbox{\kern 0.20em h}}
\def\sec{\hbox{\kern 0.20em sec}}
\def\min{\hbox {\kern 0.20em min}}
\def\smu{\hbox{\kern 0.20em s$^{-1}$}}
\def\smd{\hbox{\kern 0.20em s$^{-2}$}}
\def\an{\hbox{\kern 0.20em an}}
\def\anmu{\hbox{\kern 0.20em an$^{-1}$}}
\def\deg{\hbox{\kern 0.20em $^{\rm o}$}}
\def\yr{\hbox{\kern 0.20em yr}}
\def\yrmu{\hbox{\kern 0.20em yr$^{-1}$}}
\def\Myr{\hbox{\kern 0.20em Myr}}
\def\Mymu{\hbox{\kern 0.20em Myr$^{-1}$}}
\def\K{\hbox{\kern 0.20em K}}
\def\pcmu{\hbox{\kern 0.20em pc$^{-1}$}}
\def\pcmd{\hbox{\kern 0.20em pc$^{-2}$}}
\def\pcmt{\hbox{\kern 0.20em pc$^{-3}$}}
\def\kms{\hbox{\kern 0.20em km\kern 0.20em s$^{-1}$}}
\def\kmpd{\hbox{\kern 0.20em km$^{2}$}}
\def\kpc{\hbox{\kern 0.20em kpc}}
\def\cms{\hbox{\kern 0.20em cm\kern 0.20em s$^{-1}$}}
\def\erg{\hbox{\kern 0.20em erg}}
\def\ergs{\hbox{\kern 0.20em erg}}
\def\cmpd{\hbox{\kern 0.20em cm$^2$}}
\def\cmmd{\hbox{\kern 0.20em cm$^{-2}$}}
\def\cmms{\hbox{\kern 0.20em cm$^{-6}$}}
\def\cmpt{\hbox{\kern 0.20em cm$^3$}}
\def\cmmt{\hbox{\kern 0.20em cm$^{-3}$}}
\def\mpd{\hbox{\kern 0.20em m$^2$}}
\def\mmd{\hbox{\kern 0.20em m$^{-2}$}}
\def\mpt{\hbox{\kern 0.20em m$^3$}}
\def\mmt{\hbox{\kern 0.20em m$^{-3}$}}
\def\mujy{\hbox{\kern 0.20em $\mu$Jy}}
\def\mjy{\hbox{\kern 0.20em mJy}}
\def\Mj{\hbox{\kern 0.20em MJy}}
\def\jy{\hbox{\kern 0.20em Jy}}
\def\ghz{\hbox{\kern 0.20em GHz}}
\def\srmd{\hbox{\kern 0.20em sr$^{-1}$}}
\def \kms{km~$\rm{s}^{-1}$}

\def \mum{$\mu$m}

\def\G{\hbox{\kern 0.20em G}}

\def\h13cop{\hbox{H$^{13}$CO$^{+}$}}

\def\S+{\hbox{S{\small II}}}

\slugcomment{The Evolving ISM in the Milky Way \& Nearby Galaxies}

\shorttitle{Polluters of the ISM}
\shortauthors{Hora et al.}

\begin{document}

\newcommand{\jfourteen}{\hbox{$J=14\rightarrow 13$}}
 \title{Planetary Nebulae: Exposing the Top Polluters of the ISM}

\author{Joseph L. Hora\altaffilmark{1},
Massimo Marengo\altaffilmark{1}, Howard A. Smith\altaffilmark{1}, 
Luciano Cerrigone\altaffilmark{1}, William B. Latter\altaffilmark{2}
}
\altaffiltext{1}{Harvard-Smithsonian Center for Astrophysics}
\altaffiltext{2}{NASA/Herschel Science Center}

\begin{abstract}
The high mass loss rates of stars in their asymptotic giant branch (AGB) stage of 
evolution is one of the most important pathways for mass return from stars to the 
ISM. In the planetary nebulae (PNe) phase, the ejected material is illuminated and can 
be altered by the UV radiation from the central star. PNe therefore play a 
significant role in the ISM recycling process and in changing the environment 
around them.

We show some highlights of the results of observations that have been carried out 
using the Spitzer instruments to study the gas and dust emission from PNe in 
the Milky Way and nearby galaxies. Spitzer is especially sensitive to the cool 
dust and molecules in the PNe shell and halos. We present new results from our 
program on Galactic PNe, including IRAC and IRS observations of NGC~6720 in the 
ring and halo of that nebula.

\end{abstract}


\shortauthors{Hora et al.}

\shorttitle{Planetary Nebulae}

\section{Introduction}

A key link in the recycling of material to the Interstellar Medium (ISM) is the 
phase of stellar evolution from Asymptotic Giant Branch (AGB) to white dwarf star.
When stars are on the AGB, they begin to lose mass at a prodigious rate.  The 
stars on the AGB are relatively cool, and their atmospheres are a fertile environment 
for the formation of dust and molecules.  The material can include molecular
hydrogen (H$_2$), silicates, and carbon-rich dust.  The star is fouling its immediate 
neighborhood with these
noxious emissions.  The star is burning clean hydrogen fuel, but unlike a ``green'' hydrogen vehicle that
outputs nothing except water, the star produces ejecta of various types, some of which have
properties similar to that of soot from a gas-burning automobile (Allamandola et al. 1985).  
A significant fraction of the 
material returned to the ISM goes through the AGB - PNe pathway (Kwok 2000), making these
stars one of the major sources of pollution of the ISM.

However, these stars are not done with their stellar ejecta yet.  Before the slow, 
massive AGB wind can escape, the star begins a rapid evolution where it 
contracts and its surface temperature increases.  The star starts ejecting a less massive 
but high velocity wind that 
crashes into the existing circumstellar material, which can create a shock and a higher density
shell.  As the stellar temperature increases, the UV flux increases and it ionizes the 
gas surrounding the central star, and can excite emission from molecules, heat the dust, and
even begin to break apart the molecules and dust grains.  The objects are then visible as 
planetary nebulae, exposing their long history of spewing material into the ISM, and further
processing the ejecta.  There are even reports that the central stars of some PNe may be
engaging in nucleosynthesis for purposes of self-enrichment (e.g., Dinerstein et al. 2003), which
can be traced by monitoring the elemental abundances in the nebulae.  Clearly, we must assess
and understand the processes going on in these objects in order to understand their impact on
the ISM, and their influence on future generations of stars.

We present here results primarily from the IRAC GTO study of PNe.  See reviews by Bernard-Salas et al.
(2005, 2006) for summaries of the IRS observations of PNe, and Hora (2006a, 2006c, 2008a) 
for a review of IRAC, MIPS, and IRS results.

\section{IRAC Imaging of Planetary Nebulae}
 
\begin{figure}
\centerline{
\includegraphics[width=450pt]{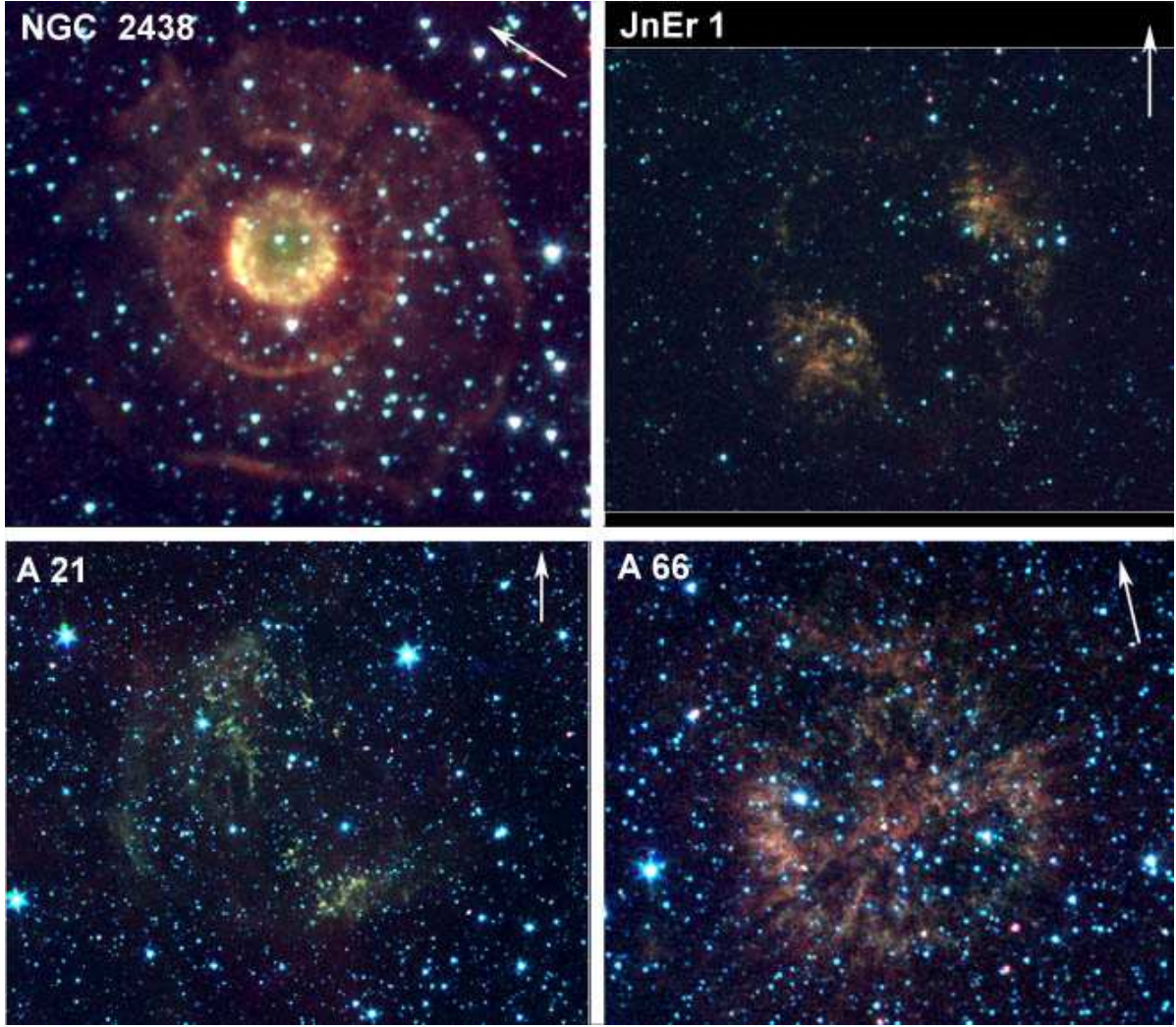}
}
\caption{\label{fig:1} Examples of PNe NGC~2438, JnEr~1, A~21, and A~66 from the IRAC GTO survey.  
The width of the images is approximately 5.5\arcmin, 9\arcmin, 14\arcmin, and 7\arcmin, respectively.
North is indicated by the direction of the arrow in the images.  
The IRAC 3.6, 4.5, and 8.0 $\mu$m bands are shown as blue, green, and red, respectively. }
\end{figure}

The IRAC camera (Fazio et al. 2004) was used to obtain images of each of the PNe at 3.6, 4.5, 5.8, and 
8.0 $\mu$m. For the GTO program on Galactic PNe (PID 68), 5-10 images per position using 30 
sec HDR frames were obtained. A total of 52 PNe were observed in the survey. More details on the
data reduction steps and some initial results were presented in Hora et al. (2004, 2006a, 2006b).
 The  IRAC images show that Spitzer can probe the faint 
extended emission from ionized gas, warm dust, PAHs, and H$_2$. The appearance of 
the extended emission in the IRAC bands is similar to the optical appearance 
in some cases, but often there are important differences. For example, in NGC~246 
an unexpected ``ring'' of emission is prominent in the longer IRAC wavelengths 
within the elliptical shell of the nebula. In PNe that are dominated by H$_2$ 
emission, such as in NGC~6720, NGC~6853, and NGC~7293, the spatial distribution 
closely matches that of the 2.12 $\mu$m H$_2$ line emission. Because the stellar emission 
and scattered light and nebular free-free continuum is much reduced at the 
IRAC wavelengths, the emission from the halo and from the dust and molecular lines 
appear more prominent.

Some new images of four PNe in the survey are shown in Figure \ref{fig:1}.  The PN NGC~2438 
appears similar to NGC~6720, with an inner bright ring of emission, and two red outer shells, with 
``spokes'' or filaments extending radially outward from the center.   In JnEr 1, the brightest part
of the nebula is in the equatorial region of the bipolar lobes, and is very clumpy and consistent
with the morphology observed in the 2.12 $\mu$m H$_2$ line.  In A~21, the emission is diffuse and 
similar to optical H$\alpha$ images.  In A~66, the brightest emission is in the same region as in optical
images, but much more clumpy and filamentary.  There is also a faint halo that extends to roughly twice 
the diameter of the optical ring.  These PNe exhibit the general characteristics of the IRAC sample 
-- there is often correspondence to the optical images, likely due to forbidden line emission in the 
IRAC bands from the ionized gas. Except for young PNe such as NGC~7027, there does not seem to be 
significant continuum emission from warm dust in the 3 - 8 $\mu$m spectral range, or any spatially 
distinct unidentified infrared (UIR) feature emission (which is 
usually associated with polycyclic aromatic hydrocarbons --
PAH).  In many PNe, emission from H$_2$ often dominates in the 
outer regions of the nebulae and the halos,
making them relatively bright at the longer wavelengths.

\begin{figure}
\centerline{
\includegraphics[width=300pt]{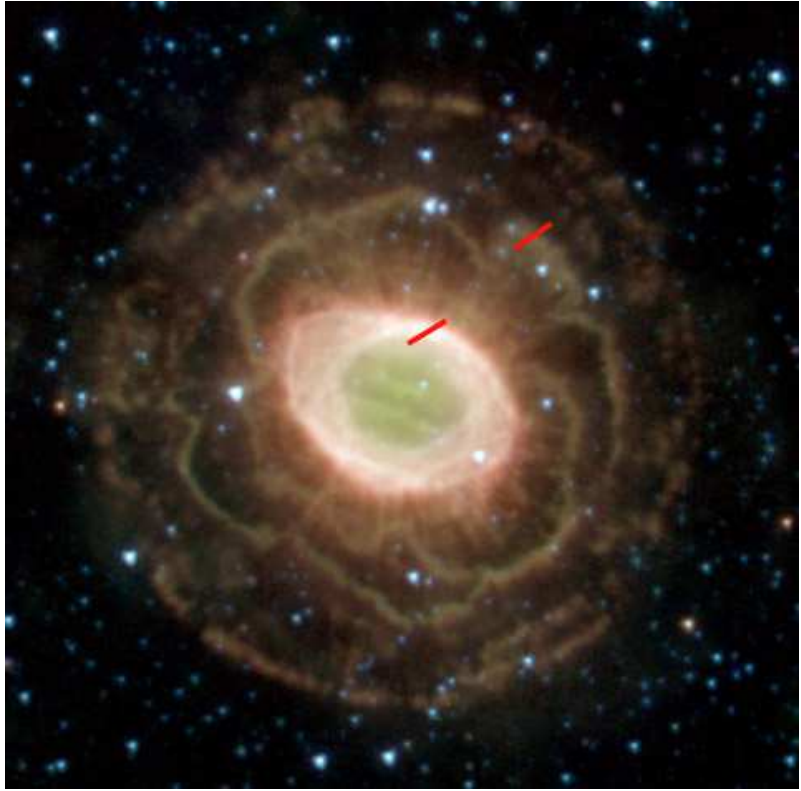}
}
\caption{\label{fig:2} The IRAC 3-color image of NGC~6720, with 3.6, 4.5, and
8.0 $\mu$m images as red, green, and blue, respectively.  North is up, and the image is 
5\arcmin~in size.  The slit positions of
the spectra shown in Figure \ref{fig:3} are shown as red boxes. }
\end{figure}

\begin{figure}
\centerline{
\includegraphics[width=250pt,angle=-90]{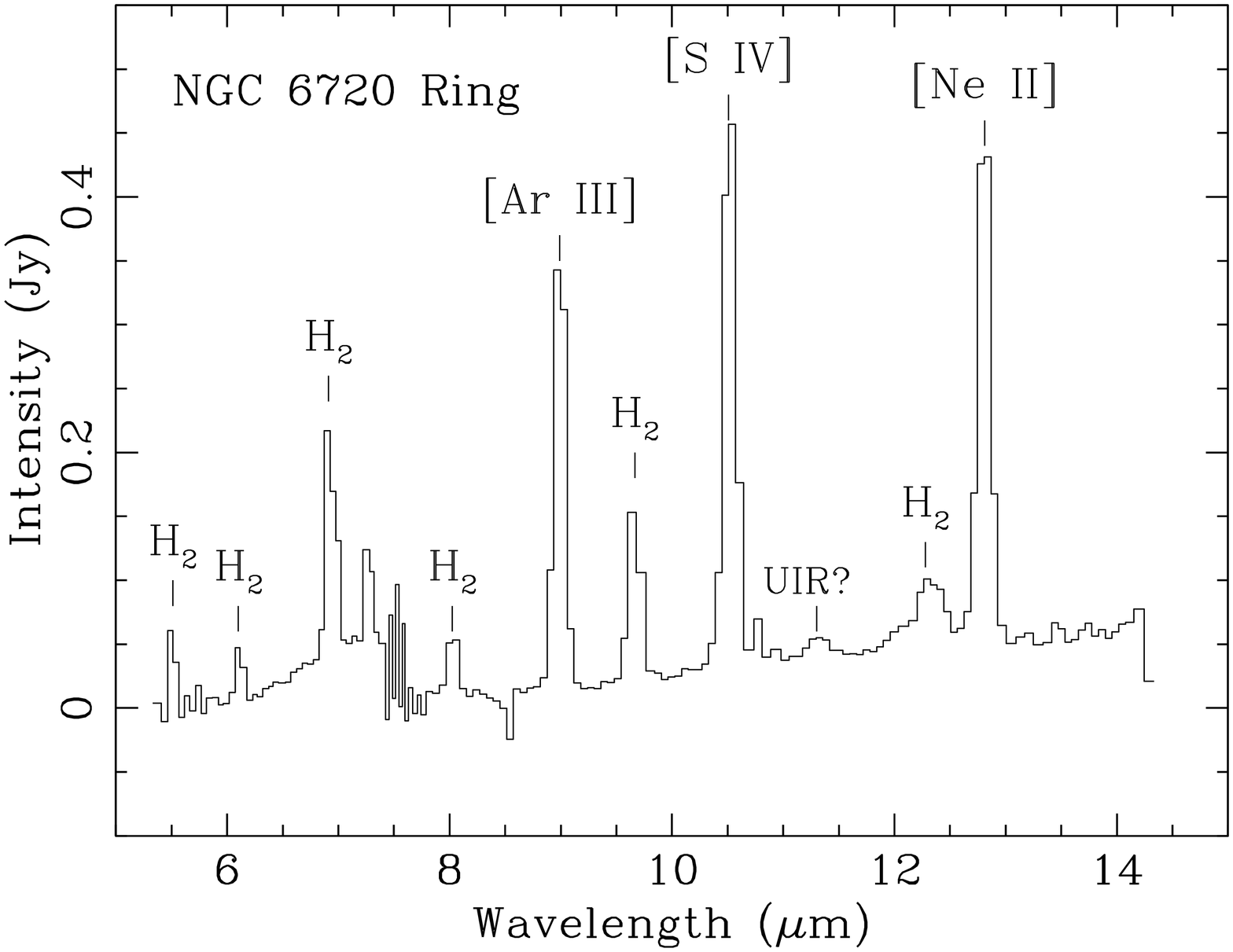}
}
\centerline{
\includegraphics[width=250pt,angle=-90]{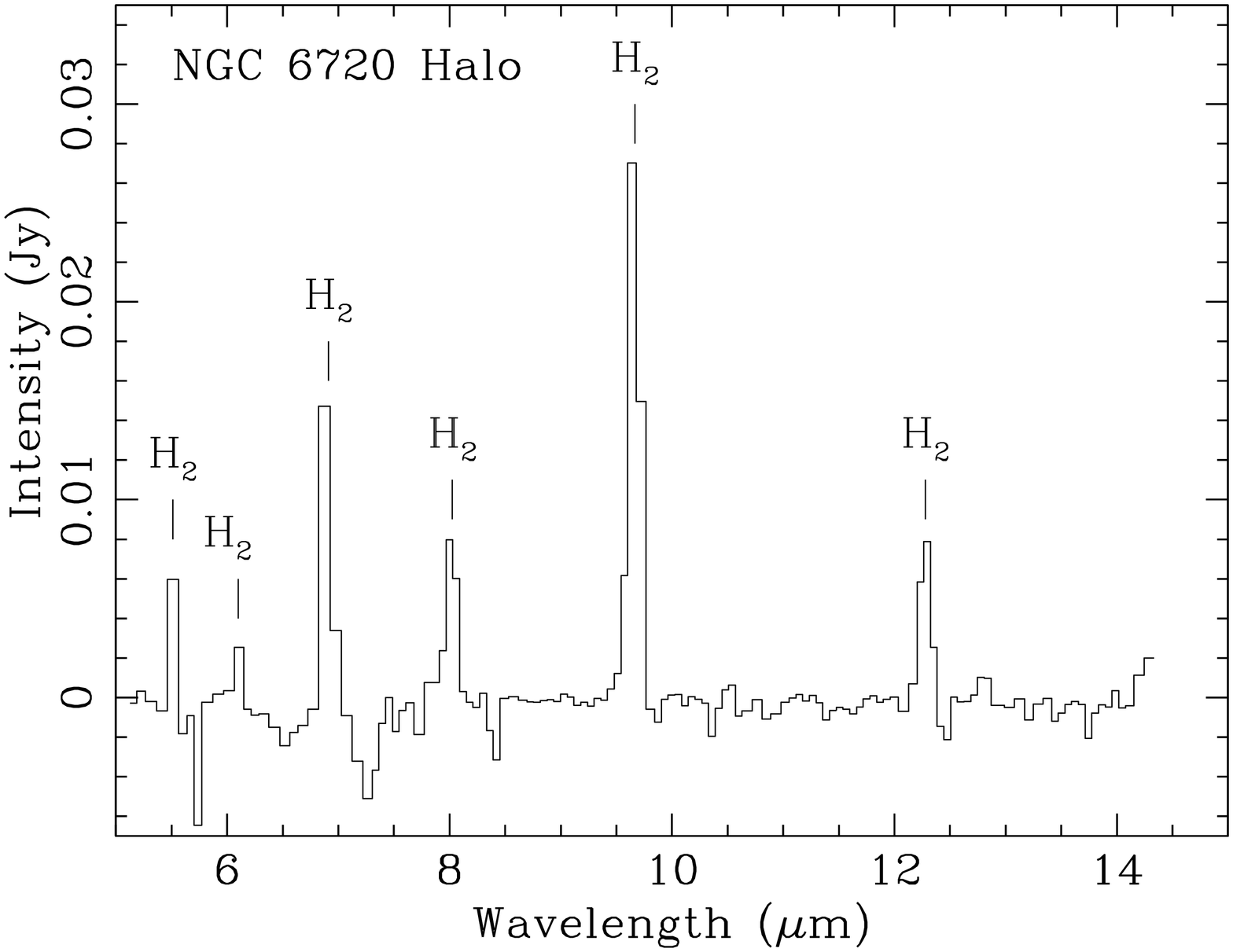}
}
\caption{\label{fig:3} The IRS low-resolution spectra of two locations in NGC~6720.  The upper spectrum
was taken at a location in the main bright ring of the nebula, and the lower spectrum
at a position in the fainter outer halo, as indicated in Figure \ref{fig:2}.  Both locations
show strong emission from H$_2$. The ``Ring'' spectrum shows forbidden line emission from
[\ion{Ar}{3}], [\ion{S}{4}], and [\ion{Ne}{2}] in the ionized gas, and 
possibly weak unidentified infrared (UIR) emission at 11.3 $\mu$m.  }
\end{figure}

An example of an object for which we have IRAC images and IRS spectra is NGC~6720.  The IRAC image
is shown in Figure \ref{fig:2}, with the IRS slit positions superposed.  The IRS low resolution 
spectra are shown in Figure \ref{fig:3}.  In the spectrum obtained on the main bright ring, the
dominant emission is from the [\ion{Ar}{3}], [\ion{S}{4}], and [\ion{Ne}{2}] lines from the ionized gas.  
However, in both spectra we see the pure rotational lines of H$_2$, similar to what is seen in NGC~7293
(Hora et al. 2006b), and in the halo these are the dominant emission feature. In the Ring spectrum, there is
some evidence for a weak broad feature near 11.3$\mu$m which could be UIR feature emission.  The 
corresponding features at 7.7 and 8.6 $\mu$m are in the overlap between the IRS orders, and have several
other features near those wavelengths, so the evidence there is not as clear.

\section{IRAC colors of PNe}
Data on the Large Magellanic Cloud (LMC) PNe were 
obtained as part of the SAGE Legacy survey (Meixner et al. 2006) and used 4$\times$12 
sec HDR frames.  The IRAC magnitudes of a set of previously known PNe in the LMC
were extracted from the SAGE data.
The IRAC [3.6] -- [4.5] vs [4.5] -- [8.0] colors of the PNe detected in all four 
IRAC bands are shown in Figure 4 (Hora et al. 2008b).  
A subset of Galactic PNe from the GTO survey and Kwok et al. (2008) study of the 
PNe in GLIMPSE are shown as green stars, the LMC PNe
are the red triangles (some of their names are labeled in blue), and the SAGE 
LMC point sources are the black dots.  The PNe are mostly in the 0.5 -- 1.2 range 
of [3.6] -- [4.5] color, and in the 1 -- 4 range of [4.5] -- [8.0] color. The PNe that 
exhibit strong PAH and/or warm dust continuum are on the right side of the plot 
(larger [4.5] -- [8.0] color), as one might expect, and have [3.6] -- [4.5] colors 
closer to 1. Strong forbidden line emission in the 8.0 $\mu$m band also results in 
objects appearing on the right side of the plot.  However, the objects with no 
continuum dust emission have lower [3.6] -- [4.5] values. The PNe with strong 
forbidden line emission and no PAH or dust continuum emission, for example SMP~83, 
are on the left side of the plot.  The Galactic and LMC PNe have similar
distributions.

\begin{figure}
\centerline{
\includegraphics[width=350pt]{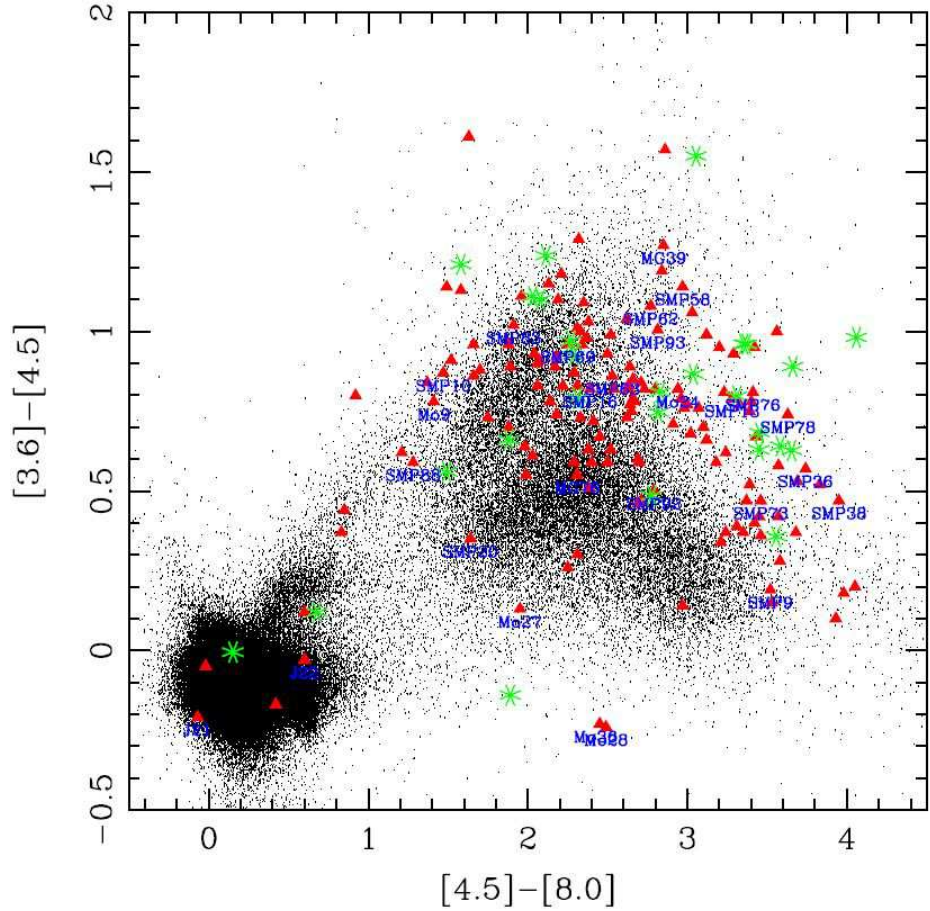}
}
\caption{\label{LMC} The LMC PNe sample (red triangles; some source names labeled in 
blue) compared to Galactic PNe
(green asterisks), from Hora et al. (2008b) and Kwok et al. (2008).  
The black points are all other 
point sources in the SAGE catalog (Meixner et al. 2006).  }
\end{figure}

\section{Summary}
The Spitzer instruments, with their unmatched sensitivity, have given us new tools to detect
and characterize the stellar ejecta that the central stars of PNe have illuminated.  Not only
will this lead to a better understanding of PNe, but will allow us to determine the properties
of the material that is being returned to the ISM.  We will then be better able to assess the
full impact of these objects on their neighbors and their surrounding environment. 

\acknowledgements

This work is based on observations made with the {\it Spitzer Space
Telescope}, which is operated by the Jet Propulsion Laboratory (JPL),
California Institute of Technology under NASA contract 1407.
Support for this work was provided by NASA and through JPL Contract 1256790.

\end{document}